\documentclass[runningheads]{llncs}
\pdfoutput=1

\usepackage{graphicx}
\usepackage{hyperref}
\usepackage{comment}
\begin{document}

\title{Choose Your Own Question: Encouraging Self-Personalization in Learning Path Construction}

\author{Anonymous Author(s)}
%
%
\author{Youngduck Choi\inst{1,2} \and
Yoonho Na\inst{1} \and
Youngjik Yoon\inst{1} \and
Jonghun	Shin\inst{1} \and \\
Chan Bae\inst{1,3} \and
Hongseok Suh\inst{1} \and
Byungsoo Kim\inst{1} \and
Jaewe Heo\inst{1}}
%
%
\institute{Riiid! AI Research \and Yale University \and UC Berkeley\\
\email{\{youngduck.choi,yh.na,youngjik.yoon,jonghun.shin,\\chan.bae,hongseok.suh,byungsoo.kim,jwheo\}@riiid.co}}
\maketitle              
\begin{abstract}
Learning Path Recommendation is the heart of adaptive learning, the educational paradigm of an Interactive Educational System (IES) providing a personalized learning experience based on the student's history of learning activities.
In typical existing IESs, the student must fully consume a recommended learning item to be provided a new recommendation.
This workflow comes with several limitations.
For example, there is no opportunity for the student to give feedback on the choice of learning items made by the IES.
Furthermore, the mechanism by which the choice is made is opaque to the student, limiting the student's ability to track their learning.
To this end, we introduce Rocket, a Tinder-like User Interface for a general class of IESs.
Rocket provides a visual representation of Artificial Intelligence (AI)-extracted features of learning materials, allowing the student to quickly decide whether the material meets their needs.
The student can choose between engaging with the material and receiving a new recommendation by swiping or tapping.
Rocket offers the following potential improvements for IES User Interfaces:
First, Rocket enhances the explainability of IES recommendations by showing students a visual summary of the meaningful AI-extracted features used in the decision-making process.
Second,  Rocket enables self-personalization of the learning experience by leveraging the students' knowledge of their own abilities and needs.
Finally, Rocket provides students with fine-grained information on their learning path, giving them an avenue to assess their own skills and track their learning progress.
We present the source code of Rocket, in which we emphasize the independence and extensibility of each component, and make it publicly available for all purposes.

\end{abstract}

\section{Introduction}
Artificial Intelligence (AI) has been applied in the field of education, allowing personalized learning paths to be generated for each individual student. 
Interactive Educational Systems (IESs) continuously use student activity data in order to provide learning materials that can most effectively benefit the student.
Despite the revolutionary impact of IESs, there has been limited innovation in their User Interfaces.
Existing IESs merely provide students with recommended learning materials and use only problem-solving data.
In particular, they by-and-large fail to collect and utilize data on student perceptions of learning materials.
By allowing students the choice of whether or not to use the recommended learning materials, we can collect data on student preferences that can inform the generation of personalized learning paths.

Another limitation of existing IESs is their opaqueness. Students are given little insight as to why a particular question was recommended for them.
This limits the trust students can have in the IES.
Furthermore, it makes it difficult for students to track the evolution of their skills.

While some IESs provide measures such as an estimated score for standardized exams, this is far from the totality of the data available to the IES for student evaluation.
By showing students the intermediate features used in recommending learning materials and estimating progress, we improve transparency and allow students to glean insights about their own progress, leading to higher levels of engagement. 

In this paper, we introduce Rocket, a Tinder-like User Interface for a general class of IESs that enable student choice in learning materials and share AI-extracted insights with the student.
Rocket provides recommendations of learning materials by showing a visual representation consisting of a personalized summary of AI-extracted insights regarding the learning material. The student may decide to engage with the material or to skip the material and indicates their choice via a swiping interface.

The benefits we expect from Rocket are as follows:

\begin{itemize}
  \item We provide an appealing visual summary of progress as the student engages with the IES. 
  \item We provide explainable learning material recommendations, giving the student insights into their learning curriculum.
  \item We provide a more personalized and more effective learning path by using student choice data in addition to traditional activity data in the AI algorithm.
  \item Students choose their own learning materials, better engaging them in the learning process and creating a more focused environment.
\end{itemize}

Finally, we publicly release the source code of Rocket with an overview of the implementation, emphasizing the modularity and extensibility of each component so that researchers and practitioners can easily use, modify, and extend our implementation.

\section{Related Works}
Rocket follows the heels of three major lines of inquiry: Tinder\footnote{\url{https://tinder.com}}-like UIs, Gamification, and Learning Path Recommendation.

Several educational services equipped with matchmaker systems have adopted Tinder-like UIs where users take swiping actions to express their personal preferences.
Future Finder\footnote{\url{https://www.salford.ac.uk/futurefinder}}, a course recommendation application at the University of Salford, allows users to swipe left or right when they are provided with personalized course recommendations. 
Papr\footnote{\url{https://jhubiostatistics.shinyapps.io/papr/}} is a web application that shows a personalized selection of papers relevant to the user's interests.
When an abstract of a paper is shown to a user, they use one of four different swiping actions (swipe up, down, right or left) to rate the paper.

Gamification is a game-based learning approach that integrates gaming features into the learning environment to enhance student motivation.
Several studies have shown that gamification contributes to improved learning outcomes by increasing learners' engagement in various learning environments, including online learning \cite{jang2015gamification} and vocational training \cite{iruela2018gamification}.
The most common gaming features include point counters, badges and leaderboards.
However, recent works have proposed more advanced designs for gamification features.
For example, \cite{monterrat2015player} presented a player model that adapts gamification features based on a learner's profile, \cite{stoeffler2018gamified} designed educational games that assess collaborative problem solving skills, and \cite{micciolo2018wearable} developed physically active multiplayer games where students not only play, but also create games.

Learning path recommendation is the task of providing personalized learning items to a student based on their history of learning activities.
Recent works have formalized the task as a reinforcement learning task and approached it equipped with virtual student simulator.
For instance, \cite{liu2019exploiting} suggested a framework for adaptive learning where an actor-critic recommender is trained to maximize cumulative rewards from a Long Short-Term Memory network-based knowledge tracing model.
Also, the deep reinforcement learning framework proposed in \cite{huang2019exploring} suggested multi-objective reward function which incorporates domain-specific factors including review, explore, smoothness and engagement.

\begin{figure*}[t]
\centering
\includegraphics[width=0.68\textwidth]{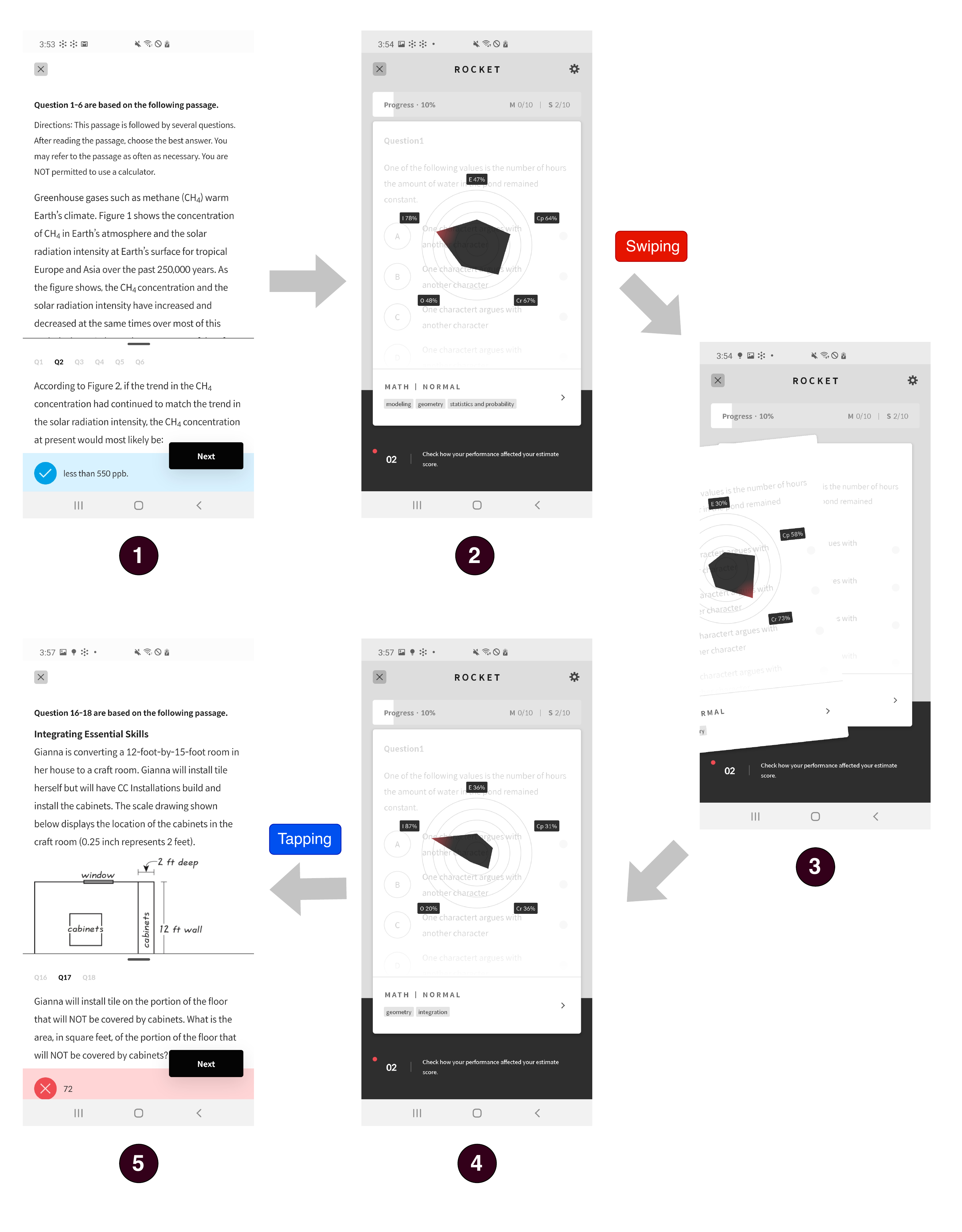}
\caption{An example flow of a student interacting with Rocket.
When the student completes the current learning material, Rocket recommends the next learning item with corresponding visual representation of AI-extracted features (1 $\rightarrow$ 2).
The student swipes to skip the presented item (2 $\rightarrow$ 3).
Rocket suggests another learning item with corresponding visual representation (3 $\rightarrow$ 4).
The student tap to choose to engage with the presented item (4 $\rightarrow$ 5).}
\label{fig:flow}
\end{figure*}

\begin{figure*}[t]
\centering
\includegraphics[width=0.6\textwidth]{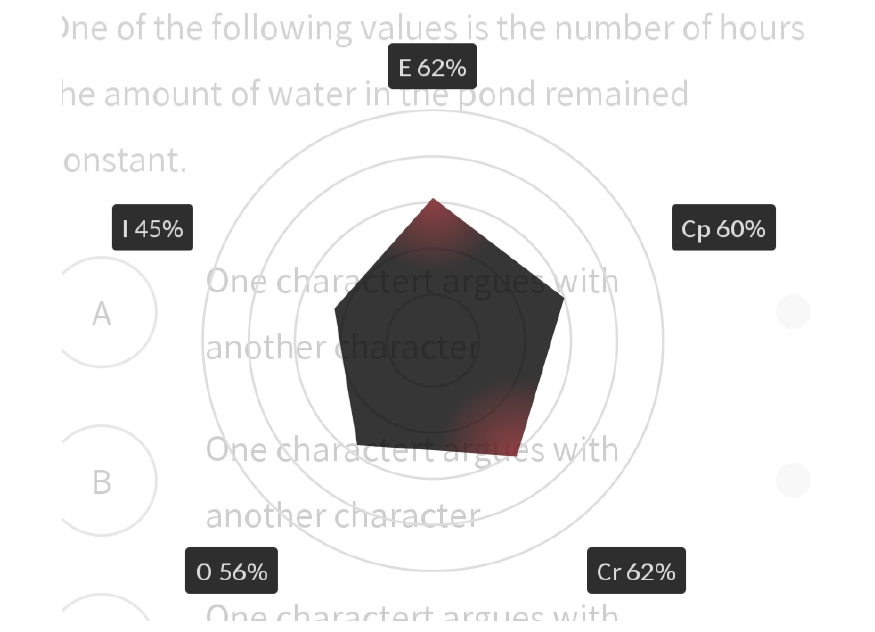}
\caption{Five AI-extracted features in pentagonal visual representation.
The features shown are expected score gain (E), completion probability (Cp), correctness probability (Cr), on-time probability (O) and initiative (I).}
\label{fig:pentagon}
\end{figure*}

\begin{figure*}[t]
\centering
\includegraphics[width=1\textwidth]{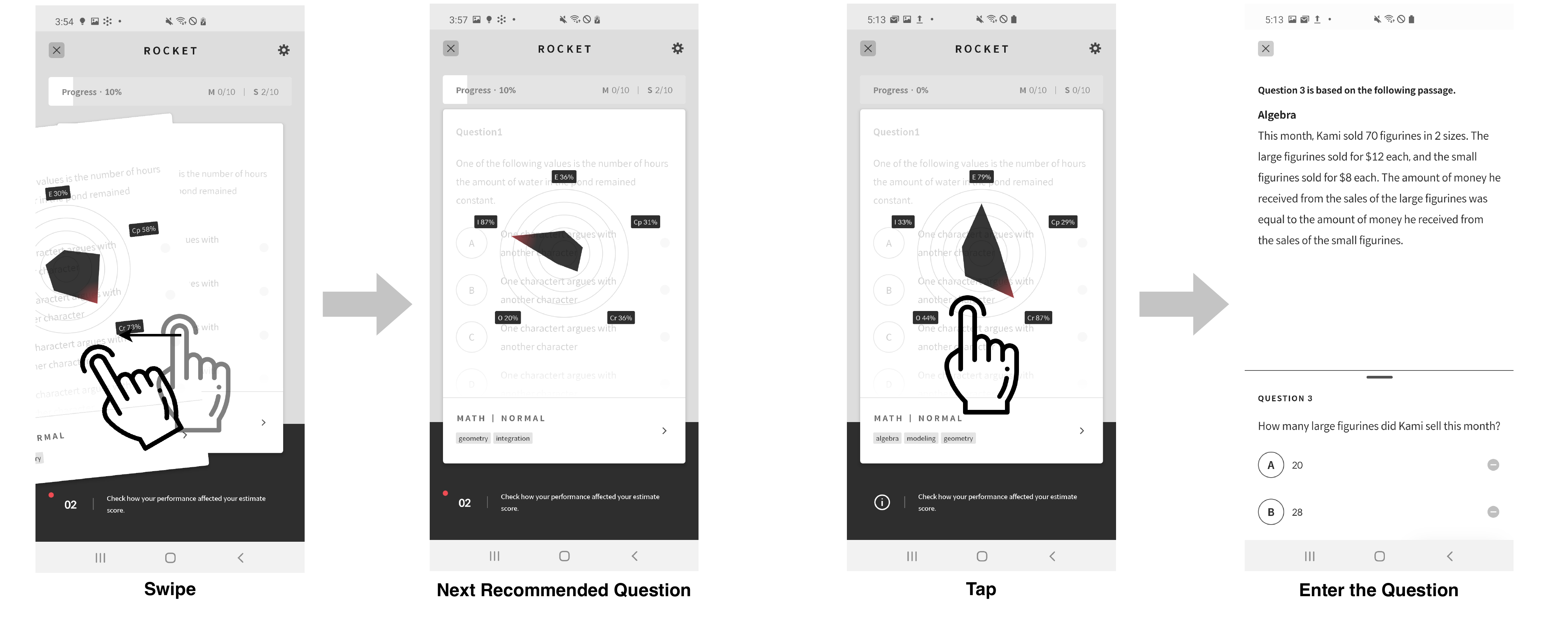}
\caption{Swiping \& tapping action on mobile devices.
Students swipe or tap to skip or engage with the presented learning item respectively.}
\label{fig:swipe}
\end{figure*}

\section{Rocket}
In this section, we describe the features of Rocket.
Rocket is a Tinder-like UI for IESs which allows students to choose between engaging with a recommended learning item and skipping this item to receive another recommendation.
For each suggested learning item, Rocket provides the student with a visual representation of AI-extracted features.
The student can then decide whether to engage with the suggested item by taking a swiping or tapping action.
Figure \ref{fig:flow} shows an example scenario of a student interacting with Rocket.
Two key features of Rocket are the polygonal visual representation of AI-extracted features and the selection UI based on swiping and tapping actions.
The following subsections describe each feature in detail.

\subsection{Visual Representation of AI-Extracted Features}
The polygonal visual representation shows features calculated by an AI-engine running behind Rocket.
The features can include analyses of the recommended contents, predicted student responses and estimated progress of the student's abilities and skills conditional on successful consumption of the recommended learning item.
Each vertex of the polygon represents an individual AI-extracted feature and the distance from each vertex to the center represents the normalized magnitude of each feature.
Figure \ref{fig:pentagon} represents an example of pentagon shaped visual representation with the five following AI-extracted features:

\begin{itemize}
  \item \textbf{Expected Score Gain (E):} Predicted increase in score if the student correctly answers the question being presented.
  Note that this is not necessarily the score value of the question since the student's response to the question will affect our model's assessment of the student's understanding of the subject domain as a whole.
  \item \textbf{Completion Probability (Cp):} The probability that the student will not quit the current learning session after consuming the question being presented.
  \item \textbf{Correctness Probability (Cr):} The probability of the student correctly answering the question being presented.
  \item \textbf{On-Time Probability (O):} The probability of the student answering the presented question within the time limit recommended by domain experts.
  \item \textbf{Initiative (I):} A measure of how different the presented question is from the previous question solved.
\end{itemize}

\subsection{Selection through Swiping or Tapping Action}
Learning item selection is done by the student taking a swiping or tapping action on a mobile touchscreen or web application (Figure \ref{fig:swipe}).
If the student swipes, the currently presented learning item is skipped and Rocket suggests another learning item.
If the student taps, it indicates that the currently presented learning item satisfies their needs and Rocket allows the student to start consuming the item.

\section{Potential Benefits of Rocket}
Students use IESs to identify their learning ability before taking the exam and to obtain a learning path that is optimized for them.
With this in mind, we expect the following benefits from introducing Rocket.

\subsection{Visual Summary of Progress}
Rocket visualizes the progress of learning ability in several ways.
The AI extracted features are provided in real-time when students access recommended learning materials.
They can monitor their abilities as they engage with many learning materials over time.
For example, the expected score gain feature is constantly updated to track their learning ability as they progress along the learning path.
In short, students can assess their own progress of learning ability.

\subsection{Explainability}
In existing IESs, the process by which learning material is recommended is completely opaque to the user.
On the other hand, Rocket shows AI extracted features reflecting the students predicted abilities.
This gives the students insight into why the learning material is being recommended, and improves the credibility of the service by showing students evidence of the AI working that they can directly assess.

\subsection{Self-Personalization}
We enable self-personalization, the process of leveraging student self-knowledge to select learning materials most appropriate to their personal situation.
As student learning preferences do not exist in a vacuum and interact with other properties of the students, we can use the provided data as input to a statistical model to predict features needed by IESs, not limited to learning preferences.
Even though IESs have access to a tremendous amount of student activity data, the insights that can be gleaned from this are decidedly limited and of an indirect nature compared to the understanding that the student has of their own abilities and needs.
As a result, they can only provide a standardized learning path that may not mesh with the students' preferred learning patterns.
For example, some students may prefer to solve a higher proportion of difficult questions to maximally challenge themselves, while others may find this demotivating.
Recording student choices in learning materials provides the information necessary to classify student learning preferences, which we can use to boost student engagement by providing a self-personalized learning path. 

\subsection{Self-Engagement}
Finally, the students playing an active role in choosing learning materials lead to deeper engagement.
Existing recommendation systems merely provide learning materials processed as recommended by an AI algorithm. 
Even though the AI algorithm is constantly receiving data created in the learning path, students play a passive role and cannot feel that they are interacting with the service. 
On the other hand, Rocket gives students the active choice of choosing between learning materials, and along with it the feeling that their learning paths are created by their choices.
Furthermore, the combination of active student choices and visible metrics that respond to student choices and performance adds a gamification aspect to Rocket. This aspect of Rocket enhances the effect.
This gives students a feeling of responsibility for their own paths and results in increased engagement as has been found in \cite{bishop2006true}.

\section{Implementation}
In this section, we give a high-level overview of implementation of Rocket.
To support a wide variety of learning materials and AI-extracted features, each component should be scalable and independent of each other, and one should be able to adjust a set of parameters to change the behavior of the components, and compare those behaviors to figure out which one is better or worse.
Hence we focused on scalability and usability while implementing Rocket in practice.
We have two components, RadarChartLayout for visualizing AI-extracted features, and CardStackView for providing the UI.

\subsection{RadarChartLayout}
RadarChart, which contains the logic for drawing polygons, forms the center of RadarChartLayout. Most libraries implementing polygonal graphs can only deal with usecases in which the data labels are the axis values for each axis.
RadarChartLayout extends RadarChart to allow the flexible customization of RadarChart.

\begin{figure*}[t]
\centering
\includegraphics[width=0.8\textwidth]{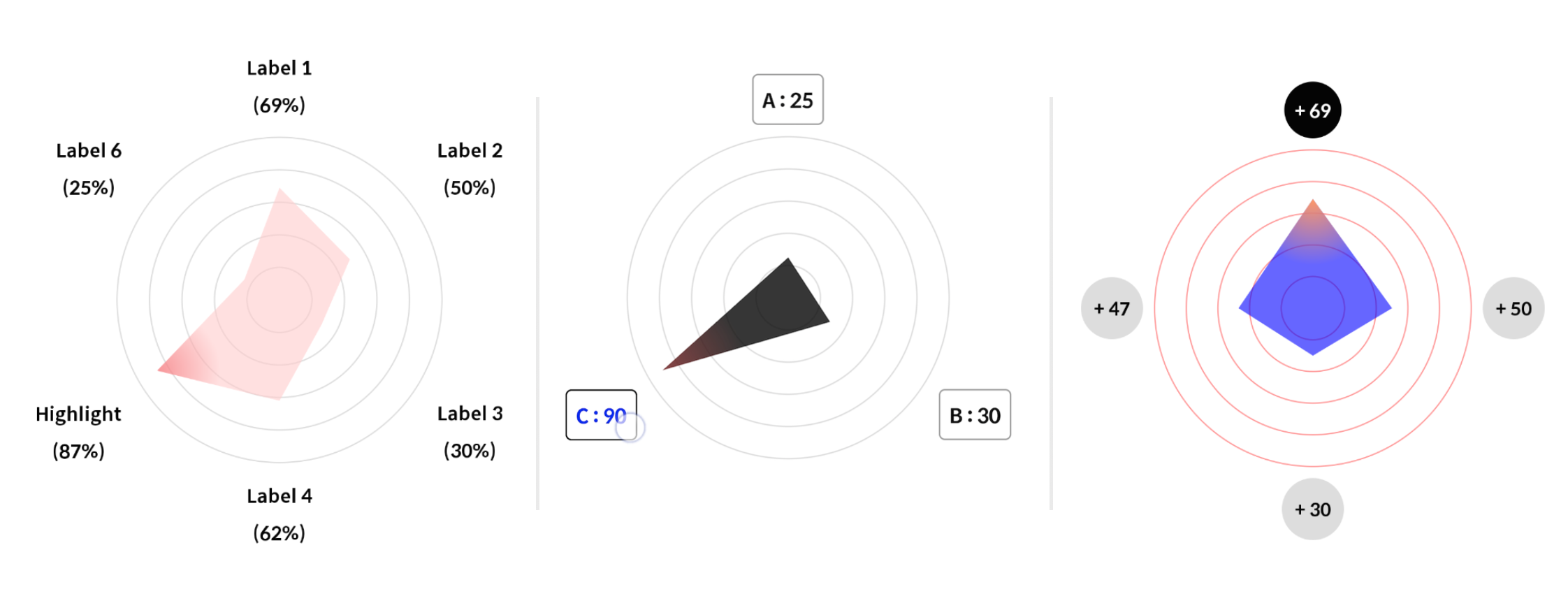}
\caption{Examples of radar charts drawn by RadarChartLayout.}
\label{fig:radar}
\end{figure*}

RadarChartLayout uses the Adapter and ViewHolder patterns to manage data and show labels.
Each Adapter manages data, and each ViewHolder defines a method to show labels.
Since Adapters and ViewHolders are independent of each other, they can be written separately and freely combined as needed.
See Figure \ref{fig:radar} for examples of radar charts combining different Adapters and ViewHolders.

\subsection{CardStackView}
CardStackView\footnote{Our implementation is a fork of \url{https://github.com/yuyakaido/CardStackView}} is the component that allows users to swipe or tap to select learning materials.
This component was implemented by with the use of RecyclerView and in particular, its LayoutManager, SmoothScroller, SnapHelper, and State. RecyclerView is a part of Android Jetpack\footnote{\url{https://developer.android.com/reference/androidx/recyclerview/widget/RecyclerView}}.
The roles of each part are as follows.

\begin{itemize}
    \item \textbf{LayoutManager:} Defines how each card will be rendered, controls card order and position, runs interactions with other components.
    \item \textbf{SmoothScroller:} Updates card position as a card is being swiped.
    \item \textbf{SnapHelper:} Controls responses (such as switching to the next card or cancelling the motion) to a card being swiped depending on swipe speed and distance.
    \item \textbf{State:} Contains the properties (location, position, action) of cards, and every component communicates with each other using the data contained in State.
\end{itemize}

\begin{figure*}[t]
\centering
\includegraphics[width=0.6\textwidth]{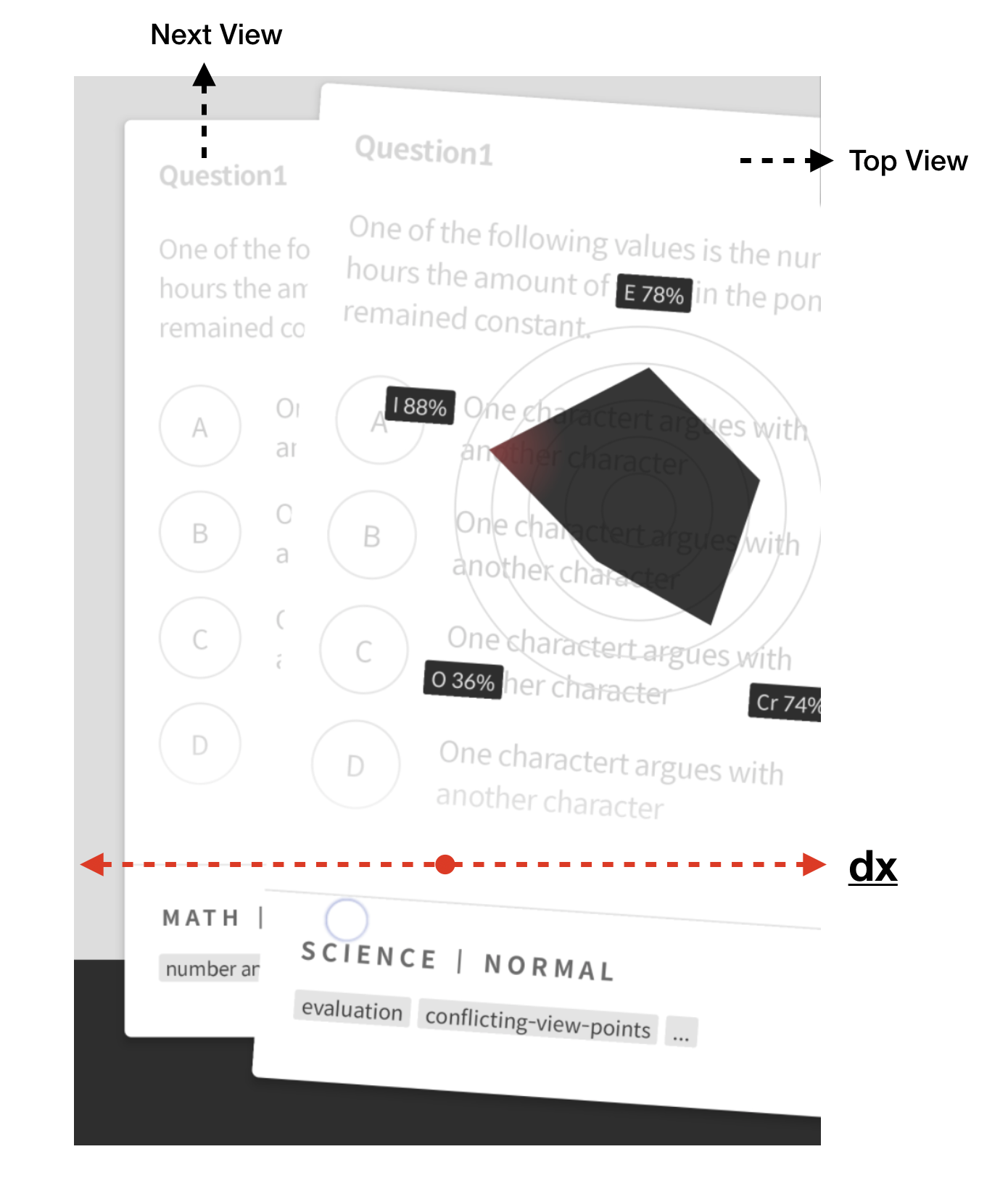}
\caption{The rendering of a card}
\label{fig:card}
\end{figure*}

There are two types of cards, a \textit{top view} currently interacting with the user, and a \textit{next view} representing the queued cards waiting to be shown.
The principal parameter controlling how a card is rendered is \textbf{dx}, the (signed) distance the card has been dragged.
It is possible to control precisely how the card moves at a given \textbf{dx}.
For instance, the card could move rightward in a rolling motion while growing bigger, or it could simply be dragged right without any size or angle changes.
The behavior of the next view is also controlled by \textbf{dx}.
In a typical scenario, the next view might grow bigger while becoming less transparent as \textbf{dx} increases.
It is straightforward to extend CardStackView to include new custom behaviors.

\begin{figure*}[t]
\centering
\includegraphics[width=0.6\textwidth]{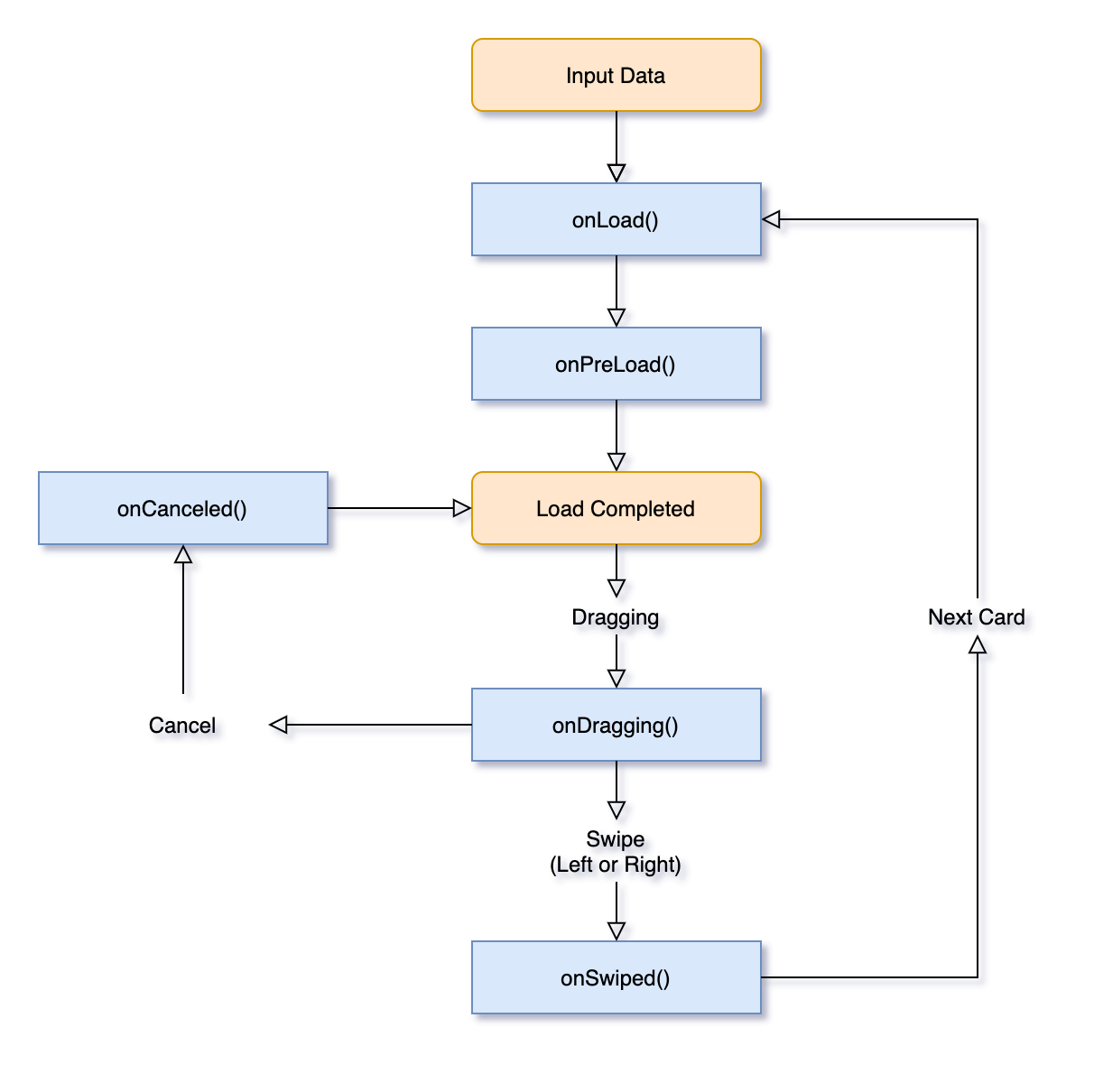}
\caption{The lifecycle of a card in CardStackView.}
\label{fig:lifecycle}
\end{figure*}

Each card has a well-defined lifecycle inside CardStackView.
The card is first loaded when information about the card is fed into the Adapter in RecyclerView.
The event that happens at this point is onLoad(), which specifies the card being loaded.
Then the onPreLoad() event returns information on the cards queued behind the current card.
After all the cards have been drawn, the user may control the cards by dragging.
While dragging, the onDragging() event tracks the direction and magnitude of the drag.
Finally, in the case the user does not drag the card and merely ceases to control the card to cancel said card, the onCanceled() event triggers.
In the other case where the user follows through with a swipe, the onSwiped() event returns information about the swiped card.
At this point, the card is considered consumed and the process repeats with the next card in the queue.

The lifecycle of a card is a natural abstraction for our purposes.
First, it allows us to easily track and categorize user activity.
For instance, it could be the case that specific demographics spend more time at some point in the UI.
Lifecycles provide a convenient framework both for collection and for making sense of this data.
Furthermore, lifecycles are well-suited for applying animations to radar charts.
When a card is shown in the top view, the associated radar chart may be animated.
But when a card is being shown in the next view as the previous card is being dragged, the card in the next view must not be animated as it may be distracting to the user whose focus should be on the top view.
The lifecycle enforces this constraint by hiding the information associated with a card until it is at the front of the queue (and therefore at the top view).

\section{Conclusion}
We presented Rocket, a Tinder-like UI for learning material selection in IESs.
We hope that Rocket enables IESs to leverage the understanding that students have of their own needs, to empower students to shape their own learning while maintaining control of the larger picture and curriculum, and be a stepping stone for further research by providing a source of data containing insights on students as extracted by the students themselves.
We further presented information on the implementation of Rocket, and hope that it can serve as a guide to other researchers and practitioners in implementing similar systems with an eye towards usability and extensibility.
We will follow up on this paper with a future paper detailing empirical experiments on the effects of Rocket on student behaviour and outcomes.

\bibliographystyle{splncs04}
\bibliography{references.bib}
    
\end{document}